# DFT based investigation of structural, elastic, optoelectronic, thermophysical and superconducting state properties of binary Mo$_3$P at different pressures

*Md. Sohel Rana, Razu Ahmed, Md. Sajidul Islam, R.S. Islam, S.H. Naqib\**
*Department of Physics, University of Rajshahi, Rajshahi 6205, Bangladesh*
*\*Corresponding author email: [salehnaqib@yahoo.com](mailto:salehnaqib@yahoo.com)*

**Abstract**

In recent years, the investigation of novel materials for various technological applications has gained much importance in materials science research. Tri-molybdenum phosphide (Mo$_3$P), a promising transition metal phosphide (TMP), has gathered significant attention due to its unique structural and electronic properties, which already make it potentially valuable system for catalytic and electronic device applications. Through an in-depth study using the density functional theory (DFT) calculations, this work aims to clarify the basic properties of the Mo$_3$P compound at different pressures. In this work, we have studied the structural, elastic, optoelectronic and thermophysical properties of binary Mo$_3$P compound. In this investigation, we varied uniform hydrostatic pressure from 0 GPa to 30 GPa. A complete geometrical optimization for structural parameters is performed and the obtained values are in good accord with the experimental values where available. It is also found that Mo$_3$P possesses very low level of elastic anisotropy, reasonably good machinability, ductile nature, relatively high Vickers hardness, high Debye temperature and high melting temperature. Thermomechanical properties indicate that the compound has potential to be used as a thermal barrier coating material. The bonding nature in Mo$_3$P has been explored. The electronic band structure shows that Mo$_3$P has no band gap and exhibits conventional metallic behavior. All of the energy dependent optical characteristics demonstrate apparent metallic behavior and agree exactly with the electronic density of states calculations. The compound has excellent reflective and absorptive properties suitable for optical applications. Pressure dependent variations of the physical properties are explored and their possible link with superconductivity has been discussed.

# 1. Introduction

In non-centrosymmetric superconductors (NCSCs), there is a lack of inversion symmetry which induces an antisymmetric spin-orbit coupling (ASOC) as a consequence of an emerging electric field gradient [1]. The admixture of spin-singlet and spin-triplet pairing in NCSCs can cause non-centrosymmetric superconductors to exhibit markedly different properties than normal superconducting systems. Despite numerous examples of NCSCs, tri-molybdenum phosphide ($Mo_3P$) reported here, may help to identify the origin of this behavior. Superconductivity in $Mo_3P$ was first predicted in 1954 [2] and confirmed ten years later [3]. In that same year, also the $Mo_3P$ crystal structure was determined [4]. Despite a non-centrosymmetric crystal structure, $Mo_3P$ is shown to be a moderately correlated electron material, which adopts a fully-gapped, spin-singlet superconducting state at low temperature. It has been found that tri-molybdenum phosphide ($Mo_3P$) nanoparticles can be a promising catalyst candidate, owing to its unique structure; its surface provides a high density of molybdenum (Mo) active sites with special electronic properties, that is, low work function and high density of d-orbital electrons at Fermi energy, which can also be promising for electrocatalysis [5]. The advancement of lithium–air (Li–air) batteries, proposed as a potential alternative for existing energy storage systems, is mainly hampered by low energy efficiency and limited life cycle. One of the biggest disadvantages of today's Li-air batteries is that developed catalysts have poor activity for both the oxygen reduction and evolution reactions (ORR and OER) [6]. Density functional theory calculations on $Mo_3P$ show that the observed strong ORR and OER electrocatalytic activity and minimal discharge/charge overpotentials in this compound are due to an oxide overlayer developed on the surface.

In this work, we provide the results of an exhaustive analysis of the physical characteristics of $Mo_3P$ using Kohn-Sham density functional theory (KS-DFT). Several physical properties of $Mo_3P$ have been studied both theoretically and experimentally so far [4,5,7-9]. But still there are notable shortages of information on this material. This lack of information limits the feasibility of potential applications. To the best of our knowledge, many important physical properties (elastic, bonding, optoelectronic and thermo-physical properties) remain unexplored. Specially, pressure dependent investigations of the physical properties have not yet been discussed at all. It is important to emphasize that any practical application of a compound requires a thorough understanding of its mechanical/elastic response to external stress. Understanding elastic constants is helpful in understanding bonding strength, solid stiffness, and mechanical stability. It is important to look into mechanical qualities (such brittleness and ductility) in order to determine whether a material is suitable for making a specific device. Elastic anisotropy indices, on the other hand, provide details on potential mechanical failure modes. When choosing a system for applications involving optoelectronic devices, optical characteristics are crucial to understand. The behavior of a material at various temperatures can be understood thanks to thermophysical characteristics. Pressure application impacts a compound's physical properties. For a better understanding of bonding character, electronic density of states, and superconductivity in materials, pressure-dependent elastic and band structure computations are essential [10,11]. A thorough understanding of the elastic, mechanical, thermophysical, bonding and optical response of this compound is necessary to unravel the potential of $Mo_3P$ for possible applications. This constitutes the primary motivation of the present study.

The rest of the article is organized in the following manner: In Section 2, we have discussed the computational procedure. Section 3 discloses the computational results and analyses. Major findings of our work are discussed and summarized in the concluding Section 4.

## 2. Computational methodology

Density functional theory (DFT), a quantum mechanical computing technique, is used to determine the electronic structure of atoms, molecules, and solids [12]. The ground state of the crystalline system can be found by solving the Kohn-Sham equation [13]. In this study, the structural, elastic, electronic and optical properties of binary $Mo_3P$ are carried out by using DFT embedded into the CASTEP (CAmbridge Serial Total Energy Package) simulation code [12]. The choice of the pseudopotential is quite important in view of optimization of the crystal structure and electronic structure. We adopted the plane wave pseudopotential method to consider the valence electrons for each atom. To treat the exchange-correlation potentials, we applied the generalized gradient approximation (GGA) of Perdew-Burke-Ernzerhof (PBE) scheme [14]. The electronic orbitals used for Mo and P to derive the valance and conduction bands are: Mo [$4d^5\ 5s^2$] and P [$3s^2\ 3p^3$]. The total energies of each cell were calculated using periodic boundary conditions. The trial wave functions were expanded using a plane wave foundation. For the structure optimization, a plane-wave cutoff energy of 550 eV was used in the calculations. The Brillouin zone (BZ) was sampled using k-point grids produced using the Monkhorst-Pack technique [15]. The structure optimization and energy calculation convergence quality was set to ultra-fine with a k-point mesh of 3 × 3 × 6. The CASTEP programme is used to calculate the independent elastic constants $C_{ij}$, bulk modulus B, and shear modulus G using the 'stress-strain' approach. The electronic band structure features were estimated using the optimized geometry of the $Mo_3P$ crystal. To determine all of the optical constants, we analyzed the photon-induced electronic transition probabilities between distinct electronic orbitals. The optical characteristics can be calculated using the complex dielectric function:

$$\varepsilon(\omega) = \varepsilon_1(\omega) + i\varepsilon_2(\omega)$$

The real part $\varepsilon_1(\omega)$ of dielectric function $\varepsilon(\omega)$ can be found from the corresponding imaginary part, $\varepsilon_2(\omega)$, which directly follows from the Kramers–Kronig relationships.

The imaginary part, $\varepsilon_2(\omega)$, is calculated within the momentum representation of matrix elements between occupied and unoccupied electronic states by employing the CASTEP supported formula expressed as:

$$\varepsilon_2(\omega) = \frac{2e^2\pi}{\Omega\varepsilon_0} \sum_{k,v,c} |<\psi_k^c|\hat{u}.\vec{r}|\psi_k^v>|^2 \delta(E_k^c - E_k^v - E) \qquad (1)$$

In this expression, $\Omega$ is the volume of the unit cell, $\omega$ is the frequency of the incident electromagnetic wave (photon), e is the electronic charge, and $\psi_k^c$ and $\psi_k^v$ are the conduction and valence band wave functions, respectively having a wave-vector k. The delta function forces energy and momentum conservation throughout the optical transition. All the other important optical parameters are estimated from the dielectric function $\varepsilon(\omega)$, once it is known at different energies using various interrelations [16-19].

The real, $n(\omega)$, and imaginary, $k(\omega)$, parts of the complex refractive index can be computed using the following relations:

$$n(\omega) = \frac{1}{\sqrt{2}}[\{\varepsilon_1(\omega)^2 + \varepsilon_2(\omega)^2\}^{1/2} + \varepsilon_1(\omega)]^{1/2} \qquad (2)$$

$$k(\omega) = \frac{1}{\sqrt{2}}[\{\varepsilon_1(\omega)^2 + \varepsilon_2(\omega)^2\}^{1/2} - \varepsilon_1(\omega)]^{1/2} \qquad (3)$$

Again, the reflectivity, $R(\omega)$, can be calculated by using the complex refractive index components:

$$R(\omega) = \left|\frac{\tilde{n}-1}{\tilde{n}+1}\right| = \frac{(n-1)^2+k^2}{(n+1)^2+k^2} \tag{4}$$

Furthermore, the absorption coefficient, $\alpha(\omega)$, the optical conductivity, $\sigma(\omega)$, and the energy loss function, $L(\omega)$, can be determined from the following equations:

$$\alpha(\omega) = \frac{4\pi k(\omega)}{\lambda} \tag{5}$$

$$\sigma(\omega) = \frac{2W_{cv}\hbar\omega}{\vec{E_0}^2} \tag{6}$$

$$L(\omega) = Im\left(-\frac{1}{\varepsilon(\omega)}\right) \tag{7}$$

In Equation (6), $W_{cv}$ is the inter-band transition probability per unit time.

The elastic characteristics are obtained from the DFT computations with the CASTEP program. The acoustic velocities in the crystals are used to compute the Debye temperature. The other thermophysical properties are calculated from the crystal density, elastic constants, and elastic moduli of $Mo_3P$.

Phonon dispersion curves (PDC) and phonon density of states (PHDOS) have been obtained applying density functional perturbation theory (DFPT) based finite displacement method (FDM) [20,21]. The phonon properties can be understood using a harmonic approximation based on the knowledge of just one fundamental quantity; the force-constant matrix. The fundamental equation regarding the force-constant matrix, $D_{\nu\mu}$, is given below:

$$D_{n\mu} = \frac{\partial^2 E}{\partial u_n \partial u_\mu} \tag{8}$$

In this equation, $E$ is the energy of the crystal and $u$ is the atomic displacement. The derivative has to be evaluated at the equilibrium atomic coordinates.

## 3. Results and Discussion

### 3.1 Structural properties

The structure of tri-molybdenum phosphide ($Mo_3P$) compound is defined by two lattice parameters, a and c. In order to determine the equilibrium geometry of the compound, the total energy is minimized at zero pressure using the generalized gradient approximation. This optimization was done with respect to the volume of the unit cell. After geometry optimization, we obtained the fully relaxed lattice parameters of $Mo_3P$ at various pressures (0, 10, 20 and 30 GPa).

The binary $Mo_3P$ compound crystallizes in the tetragonal α-$V_3S$-type structure with space group $I\bar{4}2m$ (No. 121) [7]. The unit cell of $Mo_3P$ compound is shown in Figure 1.

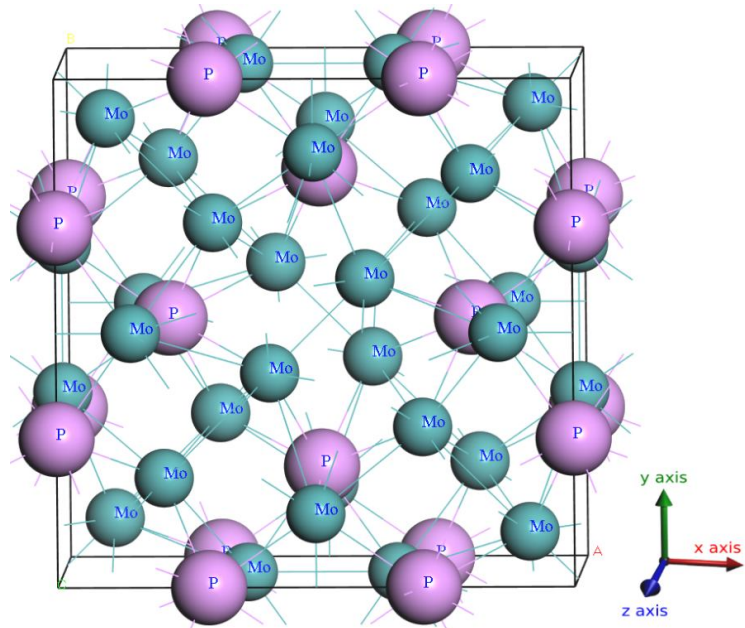

**Figure 1.** The unit cell of $Mo_3P$ crystal structure.

The unit cell of $Mo_3P$ contains 24 Mo atoms and 8 P atoms; 8 formula units in total. The optimized structural parameters of $Mo_3P$ are tabulated in Table 1 along with the experimental results for comparison [7].

**Table 1.** Calculated lattice constants a and c (both in Å), equilibrium volume $V_o$ (Å$^3$), number of formula units in the unit cell Z and total number of atoms in the unit cell N of $Mo_3P$.

| Pressure P (GPa) | a (= b) | c | c/a | $V_o$ | Z | N | Reference |
|---|---|---|---|---|---|---|---|
| 0 | 9.81 | 4.82 | 0.49 | 463.90 | 8 | 32 | This work |
| 0 | 9.79 | 4.83 | 0.49 | 462.64 | 8 | 32 | [7] |
| 10 | 9.69 | 4.77 | 0.49 | 447.59 | 8 | 32 | This work |
| 20 | 9.59 | 4.72 | 0.49 | 433.87 | 8 | 32 | This work |
| 30 | 9.50 | 4.68 | 0.49 | 422.04 | 8 | 32 | This work |

The calculated lattice parameters showed excellent agreement with the experimentally obtained values. The theoretical volume of the unit call is slightly greater than the experimental one because of the GGA, which tends to overestimate the lattice parameters slightly by spreading the electronic orbitals. On the other hand, local density approximation (LDA) tends to underestimate the lattice parameters. The difference between the theoretical and experimental values can also arise because of the experimental environment, which had a finite temperature and pressure compared to the zero temperature and pressure conditions of the theoretical one. Overall, GGA treatment results in lattice parameters very close to the experimental ones provides support about the DFT results found in this study. From Table 1, it is also seen that the lattice constants and equilibrium volume are decreasing gradually with increasing pressure, as expected.

**Table 2.** Wyckoff positions and atomic coordinates of Mo and P atoms in a unit cell of $Mo_3P$.

| Atom | Wyckoff | Occupancy | x | y | z | Reference |
|---|---|---|---|---|---|---|
| P | 8*f* | 1.0 | 0.29240 | 0.00000 | 0.0000 | [7] |
| $Mo_1$ | 8*g* | 1.0 | 0.35541 | 0.00000 | 0.5000 | |
| $Mo_2$ | 8*i* | 1.0 | 0.09289 | 0.09289 | 0.2657 | |
| $Mo_3$ | 8*i* | 1.0 | 0.29871 | 0.29871 | 0.2649 | |

The Wyckoff positions and fractional coordinates of Mo and P atoms are shown in Table 2.

**3.2 Elastic properties and anisotropy**

Elastic constants play very important role in evaluating the mechanical, lattice dynamical and many thermophysical properties of solids. Tetragonal crystals have six independent elastic constants $C_{ij}$, namely $C_{11}$, $C_{33}$, $C_{44}$, $C_{12}$, $C_{13}$ and $C_{66}$. The calculated elastic constants are presented in Table 3.

**Table 3**. Calculated elastic constants $C_{ij}$, tetragonal shear modulus $C'$ and Cauchy pressure $C''$ of Mo$_3$P (all in GPa).

| Pressure P (GPa) | $C_{11}$ | $C_{12}$ | $C_{13}$ | $C_{33}$ | $C_{44}$ | $C_{66}$ | $C'$ | $C''$ |
|---|---|---|---|---|---|---|---|---|
| 0 | 384.3 | 206.4 | 179.1 | 420.0 | 98.4 | 132.3 | 88.9 | 108.0 |
| 10 | 442.0 | 247.6 | 210.6 | 489.7 | 110.6 | 151.4 | 97.3 | 136.9 |
| 20 | 496.1 | 286.6 | 241.8 | 552.4 | 121.9 | 169.1 | 104.7 | 164.8 |
| 30 | 548.4 | 324.9 | 270.2 | 613.6 | 132.5 | 185.7 | 111.7 | 192.4 |

Elastic constants can be used to check the mechanical stability of crystals. According to Born-Huang conditions, the necessary and sufficient conditions for mechanical and elastic stability of the tetragonal crystals are [22,23]:

$$C_{11} > 0, \quad C_{33} > 0, \quad C_{44} > 0, \quad C_{66} > 0,$$

$$C_{11} > |C_{12}|, \quad 2C_{13}^2 < C_{33}(C_{11} + C_{12}), \qquad (9)$$

$$(C_{11} + C_{33} - 2C_{13}) > 0 \quad \text{and} \quad \{2(C_{11} + C_{12}) + C_{33} + 4C_{13}\} > 0$$

All the elastic constants of Mo$_3$P are positive and satisfy these mechanical and elastic stability criteria. This indicates that Mo$_3$P is mechanically stable. The same behavior follows for the pressures we have considered.

The tetragonal shear modulus, $C'$ is given by,

$$C' = \frac{(C_{11} - C_{12})}{2} \quad (10)$$

The tetragonal shear modulus is a measure of crystal's stiffness, i.e., the resistance to shear deformation by a shear stress applied in the (110) plane in the [1$\bar{1}$0] direction [24]. The value of $C'$ for Mo$_3$P at pressure 0 GPa is 88.94 GPa. The value of $C'$ increases with increasing pressure. Therefore, the stiffness of the Mo$_3$P crystal should increase with increasing pressure.

The Cauchy pressure, $C''$ is given by,

$$C'' = (C_{12} - C_{44}) \quad (11)$$

The value of $C''$ for Mo$_3$P at pressure 0 GPa is 108.05 GPa. As $C''$ is positive, Mo$_3$P should be ductile and damage tolerant. The Cauchy pressure can also be used to describe the character of chemical bonding in solids. The positive value of $C''$ indicates the presence of metallic bonding, whereas the negative value indicates the directional covalent bonding with angular character [25,26]. Therefore, the positive value of $C''$ in Mo$_3$P indicates the presence of metallic bonding. High positive value of the Cauchy pressure means high ductility and the presence of significant central force dominated atomic bondings. As the value of $C''$ increases with increasing pressure, we can say that Mo$_3$P contains more metallic bonds at higher pressures.

The isotropic bulk modulus, B and shear modulus, G (by the Voigt-Reuss-Hill (VRH) method) and Young's modulus, Y are calculated using the following well known equations [27-29]:

$$B \equiv B_H = \frac{B_V + B_R}{2} \quad (12)$$

$$G \equiv G_H = \frac{G_V + G_R}{2} \quad (13)$$

$$Y = \frac{9BG}{(3B + G)} \quad (14)$$

The shear modulus G of a material characterizes its reaction to shear stress. A large value of G indicates strong directional bonding between the atoms [30]. The bulk modulus B describes a material's response to uniform hydrostatic pressure [30]. For Mo$_3$P, the smaller value of G compared to B (Table 4) indicates that the mechanical stability will be dominated by the shear deformation. The Young's modulus Y describes the material's strain response to uniaxial stress in the direction of this stress. The covalent nature of a material increases with Young's modulus [31]. The Young's modulus provides a measure of the stiffness and thermal shock resistance of a solid. From Table 4, it is clearly noticed that the values of elastic moduli increase gradually with increasing pressure.

**Table 4.** Bulk modulus B, shear modulus G and Young's modulus Y of $Mo_3P$ (all in GPa).

| Pressure P (GPa) | $B_V$ | $B_R$ | B | $G_V$ | $G_R$ | G | Y |
|---|---|---|---|---|---|---|---|
| 0 | 257.55 | 257.52 | 257.54 | 107.41 | 105.19 | 106.30 | 280.33 |
| 10 | 301.24 | 301.21 | 301.23 | 121.50 | 118.38 | 119.94 | 317.66 |
| 20 | 342.78 | 342.75 | 342.76 | 134.19 | 130.23 | 132.21 | 351.44 |
| 30 | 382.35 | 382.33 | 382.34 | 146.49 | 141.60 | 144.04 | 383.92 |

High value of Y for $Mo_3P$ (as a comparison, the Young's modulus of steel is ~200 GPa) suggests that the material under study is fairly stiff and should show high thermal shock resistance.

The Pugh's indicator or Pugh's ratio, $G/B$, which is the ratio of shear modulus and bulk modulus, is an indicator of failure mode (ductile/brittle) of a material. If the value of $G/B$ is less than 0.57, the material is ductile, otherwise it would be brittle [32-34]. For $Mo_3P$, the value of $G/B$ at pressure 0 GPa is 0.41 (Table 5). Hence, $Mo_3P$ shows ductile nature, which once again indicates that the metallic bonding dominates over covalent bonding.

The Kleinman parameter or internal strain parameter, ξ is calculated using the following equation [35]:

$$\xi = \frac{C_{11} + 8C_{12}}{7C_{11} + 2C_{12}} \tag{15}$$

The Kleinman parameter, ξ is an indicator which measures the stability of a compound against stretching and bending. It is a dimensionless parameter whose value generally lies in the range 0 ≤ ξ ≤ 1. Bond stretching dominates for the lower limit of ξ (ξ close to zero) and bond bending dominates for the upper limit of ξ (ξ close to 1) [24]. For $Mo_3P$, the estimated value of ξ at 0

GPa is 0.66 (Table 5). This indicates that the limiting mechanical strength of $Mo_3P$ is mainly affected by bond bending contribution over bond stretching or contracting.

Poisson's ratio, $\sigma$ is given by [36],

$$\sigma = \frac{(3B - 2G)}{2(3B + G)} \tag{16}$$

For central-forces, the lower and upper limits of Poisson's ratio for solids are 0.25 and 0.50, respectively [37,38]. For $Mo_3P$, the value of $\sigma$ at 0 GPa is 0.32 (Table 5). This implies that interatomic forces are central in nature. The Poisson's ratio can also be used to determine whether a substance has ionic and covalent bonds. For ionic solids, the value of $\sigma$ is typically 0.25 and for covalent materials, the value of $\sigma$ is typically 0.10 [39]. This yardstick shows that ionic contribution is present in $Mo_3P$.

The machinability index, $\mu_M$ of a material is calculated using following equation [40]:

$$\mu_M = \frac{B}{C_{44}} \tag{17}$$

Machinability is related to a number of mechanical variables connected to the inherent properties of the work materials, cutting tool material, cutting conditions, type of cutting, cutting fluid, tool geometry, and the nature of tool engagement with the work and machine tool rigidity and its capacity. Machinability can also be used as a measure of plasticity and lubricating property of a material. Larger value of machinability of a compound indicates excellent dry lubricating properties, lower feed forces, lower friction value and higher plastic strain value [24]. The value of machinability for $Mo_3P$ at 0 GPa is 2.62 (Table 5). This value is high and comparable to those of many prominent engineering materials. From Table 5, it is seen that the value of machinability index increases with increasing pressure. This implies that the $Mo_3P$ compound shows excellent lubricating properties, lower feed forces, lower friction value and higher plastic strain value at high pressure.

The Vickers hardness, $H_v$ or micro hardness, $H_{micro}$ is given by [41],

$$H_v \equiv H_{micro} = \frac{(1-2\sigma)Y}{6(1+\sigma)} \tag{18}$$

The hardness of a material is essential to understand elastic and plastic properties [42] under high mechanical stress. The value of hardness for $Mo_3P$ at pressure 0 GPa is 12.74 GPa and increases with increasing pressure as shown in Table 5. This value of hardness implies that $Mo_3P$ is a reasonably hard material.

The macro hardness, $H_{macro}$ is calculated by [43],

$$H_{macro} = 2\left[\left(\frac{G}{B}\right)^2 G\right]^{0.585} - 3 \qquad (19)$$

The value of macro hardness for $Mo_3P$ at pressure 0 GPa is 7.89 GPa and also increases with increasing pressure (Table 5).

The fracture toughness, $K_{IC}$ is calculated using the following empirical formula [44]:

$$K_{IC} = V_o^{1/6}\sqrt{GB} \qquad (20)$$

Fracture toughness carries significant importance like the hardness for practical applications of solids. Fracture toughness gives a measure of the resistance of solids to stop the propagation of the induced fracture inside. High value of fracture toughness is expected for hard materials for use in industrial purposes [45]. The value of $K_{IC}$ for $Mo_3P$ at 0 GPa is 4.60 MPa.m$^{1/2}$. This value increases with increasing pressure.

**Table 5.** Pugh's ratio $G/B$, Kleinman parameter ξ, Poisson's ratio $\sigma$, machinability index $\mu_M$, micro hardness $H_{micro}$ or Vicker's hardness $H_v$ (GPa), macro hardness $H_{macro}$ (GPa) and fracture toughness $K_{IC}$ (MPa.m$^{1/2}$) of $Mo_3P$.

| Pressure P (GPa) | $G/B$ | ξ | $\sigma$ | $\mu_M$ | $H_v$ or $H_{micro}$ | $H_{macro}$ | $K_{IC}$ |
|---|---|---|---|---|---|---|---|
| 0 | 0.41 | 0.66 | 0.32 | 2.62 | 12.74 | 7.89 | 4.60 |
| 10 | 0.40 | 0.67 | 0.32 | 2.72 | 14.05 | 8.20 | 5.26 |
| 20 | 0.39 | 0.69 | 0.33 | 2.81 | 15.06 | 8.43 | 5.86 |
| 30 | 0.38 | 0.70 | 0.33 | 2.88 | 16.07 | 8.69 | 6.43 |

Elastic anisotropy explains the directional dependence of mechanical properties of a compound. Knowledge of elastic anisotropy is crucial for material design because elastic anisotropy is associated with lattice distortion, plastic deformation and formation of micro-cracks and mechanical failures of solids. The calculated elastic anisotropy factors are tabulated in Table 6.

**Table 6.** Shear anisotropy factors ($A_1$, $A_2$ and $A_3$), Zener anisotropy factor A, universal anisotropy index $A^U$, equivalent Zener anisotropy measure $A^{eq}$, anisotropy in compressibility $A^B$ (%), anisotropy in shear $A^G$ (%) and universal log-Euclidean anisotropy index $A^L$ of $Mo_3P$ at different pressures.

| Pressure P (GPa) | $A_1$ | $A_2$ | $A_3$ | A | $A^U$ | $A^{eq}$ | $A^B$ | $A^G$ | $A^L$ |
|---|---|---|---|---|---|---|---|---|---|
| 0 | 0.88 | 0.88 | 1.49 | 1.11 | 0.11 | 1.35 | 0.00 | 1.05 | 0.02 |
| 10 | 0.87 | 0.87 | 1.56 | 1.14 | 0.13 | 1.39 | 0.01 | 1.30 | 0.03 |
| 20 | 0.86 | 0.86 | 1.61 | 1.16 | 0.15 | 1.43 | 0.00 | 1.50 | 0.04 |
| 30 | 0.85 | 0.85 | 1.66 | 1.19 | 0.17 | 1.46 | 0.00 | 1.70 | 0.05 |

Shear anisotropy factors are used to understand the degree of anisotropy in the bonding between atoms in different planes within a crystal. The shear anisotropy for a tetragonal crystal can be quantified by three different factors [46,47].

The shear anisotropy factor for {100} shear planes between the ⟨011⟩ and ⟨010⟩ directions is,

$$A_1 = \frac{4C_{44}}{C_{11}+C_{33}-2C_{13}} \qquad (21)$$

The shear anisotropy factor for {010} shear planes between the ⟨101⟩ and ⟨001⟩ directions is,

$$A_2 = \frac{4C_{55}}{C_{22}+C_{33}-2C_{23}} \qquad (22)$$

The shear anisotropy factor for {001} shear planes between the ⟨110⟩ and ⟨010⟩ directions is,

$$A_3 = \frac{4C_{66}}{C_{11}+C_{22}-2C_{12}} \qquad (23)$$

Zener anisotropy factor, A has been calculated using the following equation,

$$A_{Zener} \equiv A = \frac{2C_{44}}{C_{11}-2C_{12}} \qquad (24)$$

The materials having isotropy in the bondings existing between different atomic planes have unit values of $A_1$, $A_2$ and $A_3$ ($A_1 = A_2 = A_3 = 1$) and any other value (greater or lesser than unity) implies that there is anisotropy in the material. The values of $A_1$, $A_2$, $A_3$ and $A$ for different pressures are given in Table 6. It is found that $Mo_3P$ shows anisotropic behavior. The degree of anisotropy increases slowly with increasing pressure.

The universal anisotropy indices $A^U$ and $d_E$, equivalent Zener anisotropy measure $A^{eq}$, anisotropy in compressibility $A^B$ and anisotropy in shear $A^G$ for a crystal with any symmetry are calculated using following standard equations [48-53]:

$$A^U = 5\frac{G_V}{G_R} + \frac{B_V}{B_R} - 6 \geq 0 \quad \text{and} \quad d_E = \sqrt{A^U + 6} \qquad (25)$$

$$A^{eq} = \left(1 + \frac{5}{12}A^U\right) + \sqrt{(1 + \frac{5}{12}A^U)^2 - 1} \qquad (26)$$

$$A^B = \frac{B_V - B_R}{B_V + B_R} \quad \text{and} \quad A^G = \frac{G_V - G_R}{G_V + G_R} \qquad (27)$$

The universal log-Euclidean index is defined as [49]:

$$A^L = \sqrt{\left\{ln\left(\frac{B_V}{B_R}\right)\right\}^2 + 5\left\{ln\left(\frac{C_{44}^V}{C_{44}^R}\right)\right\}^2} \qquad (28)$$

where, the Reuss and Voigt values of $C_{44}$ are defined as follows,

$$C_{44}^R = \frac{5}{3}\frac{C_{44}(C_{11}-C_{12})}{3(C_{11}-C_{12})+4C_{44}} \qquad (29)$$

and,

$$C_{44}^V = C_{44}^R + \frac{3}{5}\frac{(C_{11}-C_{12}-2C_{44})^2}{3(C_{11}-C_{12})+4C_{44}} \qquad (30)$$

The expression for $A^L$ is valid for all crystal symmetries. The universal anisotropy index, $A^U$ is closely related to $A^L$. For perfectly anisotropic crystal, $A^L = 0$. The values of $A^L$ range between 0 and 10.26. Almost 90% of crystalline solids have $A^L < 1$. It is also argued that, $A^L$ indicates whether the compound is layered/lamellar type [51]. The compounds that have high value of $A^L$ indicate that they have strongly layered whereas the compounds with low value of $A^L$ indicate that they have non-layered structural features. Since the calculated value of $A^L$ as shown in Table 6 is only 0.02 at 0 GPa; which is very much lower than unity, we thus conclude that Mo$_3$P will exhibit non-layered type of structural properties.

**Table 7.** Uniaxial bulk modulus $B_a$, $B_b$ and $B_c$ (GPa), isotropic bulk modulus $B_{relax}$ (GPa), anisotropies of the bulk modulus $A_{Ba}$ and $A_{Bc}$, compressibility $\beta$ (TPa$^{-1}$), linear compressibilities $\beta_a$ and $\beta_b$ (TPa$^{-1}$) and their ratio ($\beta_a/\beta_b$) for Mo$_3$P at different pressures.

| Pressure P (GPa) | $B_a$ | $B_b$ | $B_c$ | $B_{Relax}$ | $A_{Ba}$ | $A_{Bc}$ | $\beta$ | $\beta_a$ | $\beta_c$ | $\beta_c/\beta_a$ |
|---|---|---|---|---|---|---|---|---|---|---|
| 0 | 764 | 764 | 790 | 257 | 1.0 | 1.04 | 3.9 | 1.3 | 1.3 | 0.97 |
| 10 | 892 | 892 | 928 | 301 | 1.0 | 1.04 | 3.3 | 1.1 | 1.1 | 0.96 |
| 20 | 1016 | 1015 | 1054 | 342 | 1.0 | 1.04 | 2.9 | 1.0 | 1.0 | 0.96 |
| 30 | 1135 | 1135 | 1171 | 382 | 1.0 | 1.03 | 2.6 | 0.9 | 0.8 | 0.97 |

Uniaxial bulk modulus $B_a$, $B_b$ and $B_c$ (along a-, b- and c-axis, respectively) and anisotropies in the bulk modulus are calculated from the following equations [54]:

$$B_a = a \frac{dP}{da} = \frac{\Lambda}{1+\alpha+\beta} \tag{31}$$

$$B_b = a \frac{dP}{db} = \frac{B_a}{\alpha} \text{ and } B_c = c \frac{dP}{dc} = \frac{B_a}{\beta}, \tag{32}$$

$$A_{Ba} = \frac{B_a}{B_b} = \alpha \text{ and } A_{Bc} = \frac{B_c}{B_b} = \frac{\alpha}{\beta} \tag{33}$$

where,

$$\Lambda = C_{11} + 2C_{12}\alpha + C_{22}\alpha^2 + 2C_{13}\beta + C_{33}\beta^2 + 2C_{23}\alpha\beta, \tag{34}$$

$$\alpha = \frac{(C_{11}-C_{12})(C_{33}-C_{13})-(C_{11}-C_{13})(C_{23}-C_{13})}{(C_{22}-C_{12})(C_{33}-C_{13})-(C_{12}-C_{23})(C_{13}-C_{23})}, \tag{35}$$

and,

$$\beta = \frac{(C_{22}-C_{12})(C_{11}-C_{13})-(C_{11}-C_{12})(C_{23}-C_{12})}{(C_{22}-C_{12})(C_{33}-C_{13})-(C_{12}-C_{23})(C_{13}-C_{23})} \tag{36}$$

Here, $A_{B_a}$ and $A_{B_c}$ denote the anisotropies in bulk modulus along a-axis and c-axis with respect to b-axis, respectively. For Mo$_3$P,

$$A_{B_a} = \alpha = 1 \; ; \; A_{B_c} = \frac{\alpha}{\beta} = 1.04$$

Also, isotropic bulk modulus, $B_{relax}$ is given by,

$$B_{relax} = \frac{\Lambda}{(1+\alpha+\beta)^2} \tag{37}$$

The compressibility, β is given by,

$$\beta = \frac{1}{B} \tag{38}$$

The value of β is 0.00388 /GPa or 3.88 TPa$^{-1}$ at 0 GPa. The compressibility of Mo$_3$P decreases significantly with increasing pressure.

The linear compressibility of a tetragonal compound along a and c axis (β$_a$ and β$_c$) are calculated using the relations,

$$\beta_a = \frac{C_{33}-C_{13}}{D} \text{ and } \beta_c = \frac{C_{11}+C_{12}-2C_{13}}{D} \tag{39}$$

where, $D = (C_{11} + C_{12})C_{33} - 2C_{13}^2$.

The ratio of the linear compressibilities $(\beta_c/\beta_a)$ for Mo$_3$P is 0.97 at 0 GPa. From Table 7, it is seen that the value of the ratio of the linear compressibilities $(\beta_c/\beta_a)$ remains almost invariant with applying pressure.

ELATE [55] generated plots of Young's modulus, linear compressibility, shear modulus and Poisson's ratio are given below:

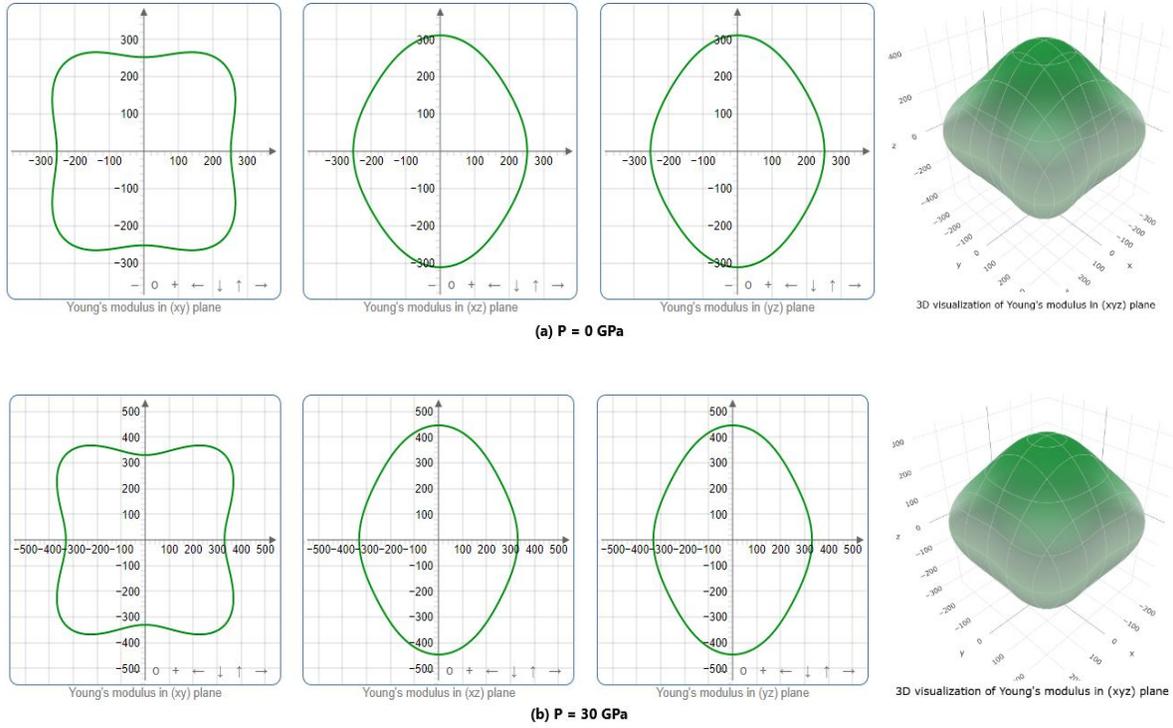

**Figure 2.** ELATE generated plots of Young's modulus at (a) P = 0 GPa and (b) P = 30 GPa.

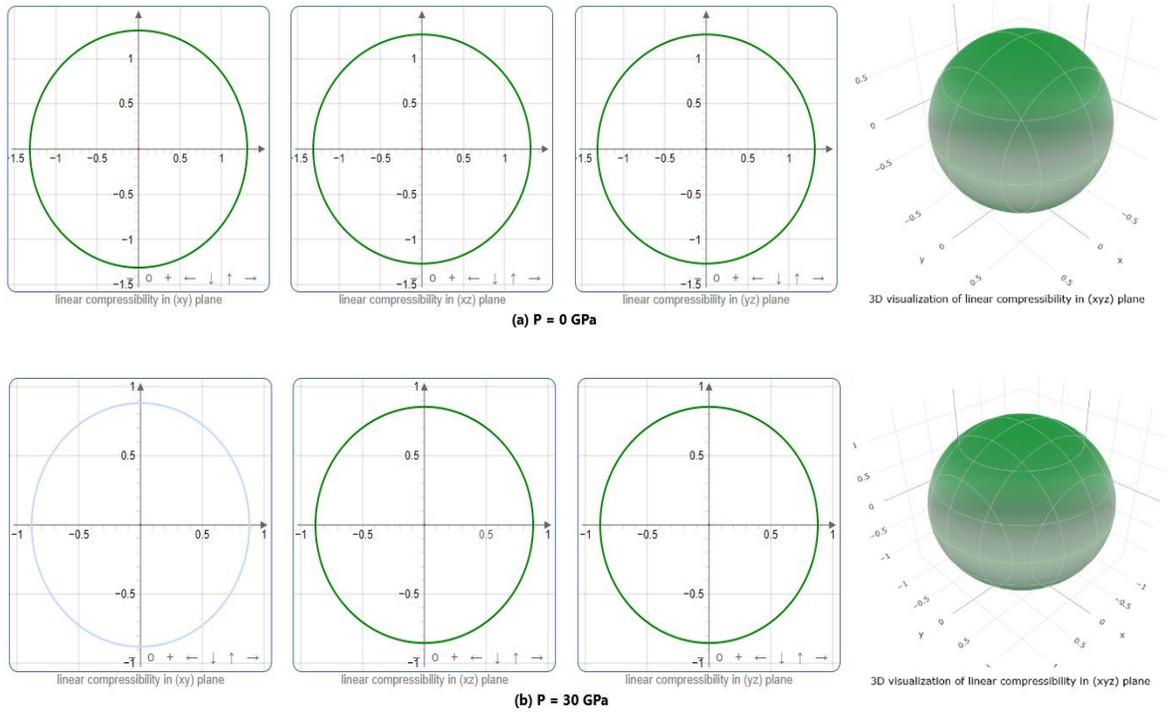

**Figure 3.** ELATE generated plots of linear compressibility at (a) P = 0 GPa and (b) P = 30 GPa.

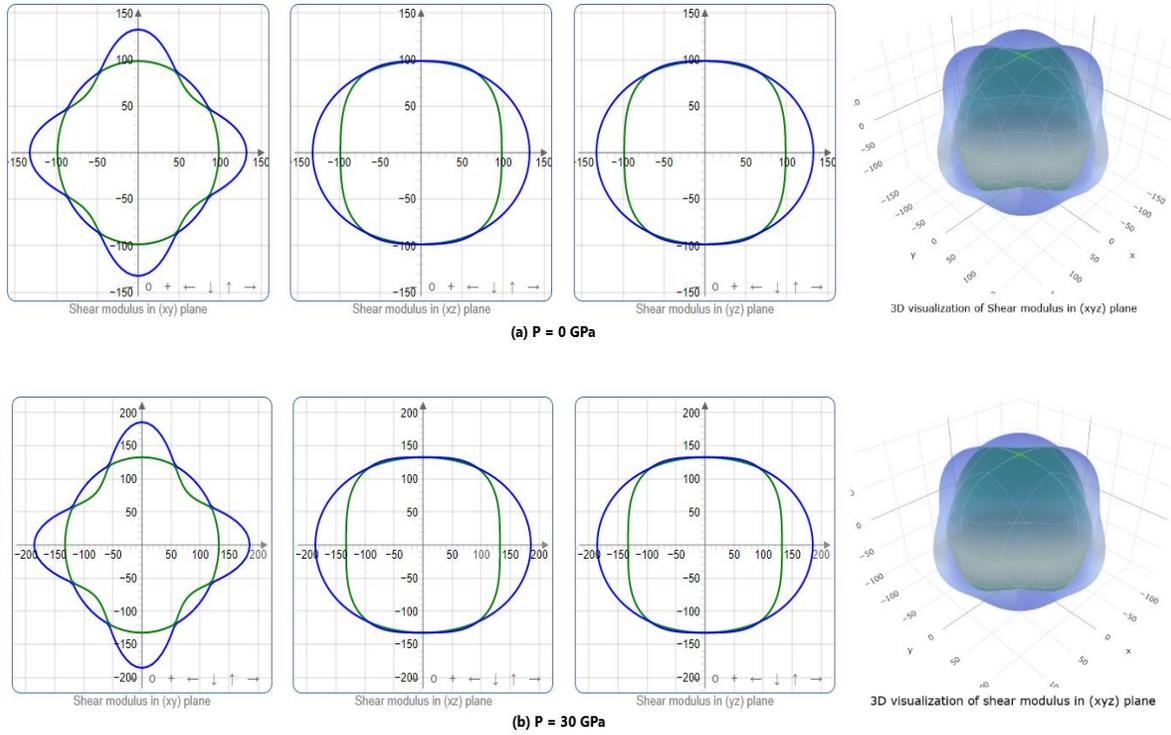

**Figure 4.** ELATE generated plots of Shear modulus at (a) P = 0 GPa and (b) P = 30 GPa.

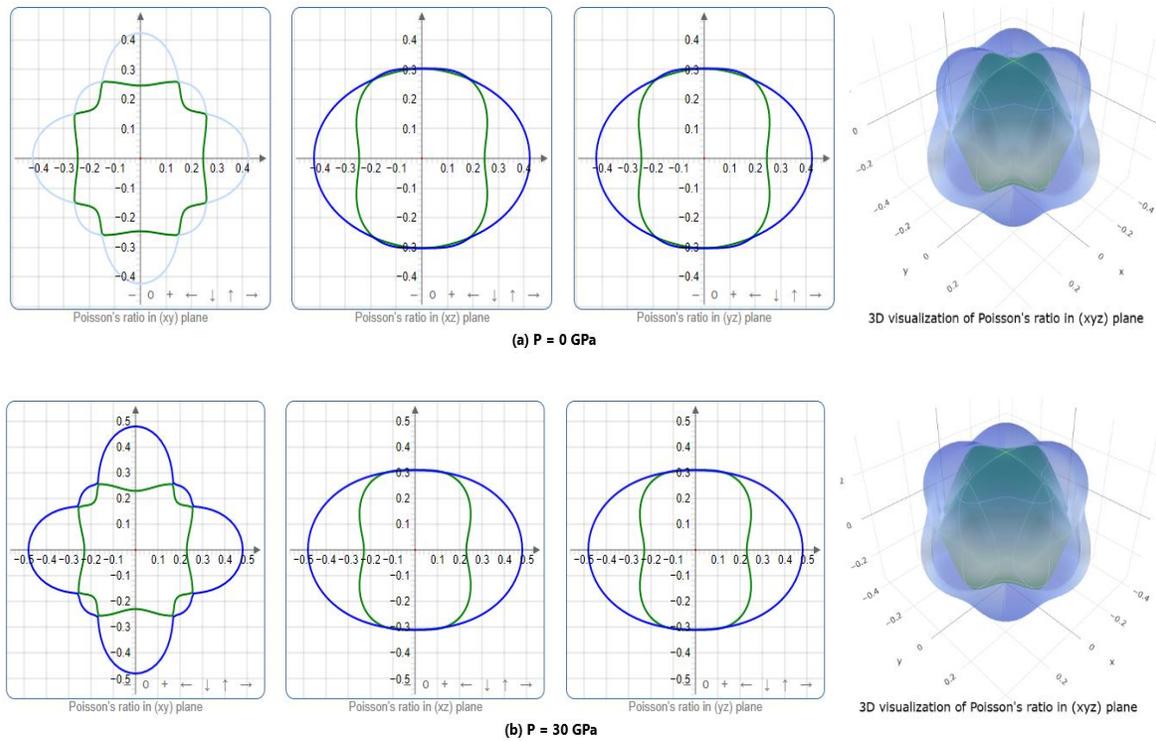

**Figure 5.** ELATE generated plots of Poisson's ratio at (a) P = 0 GPa and (b) P = 30 GPa.

For elastically isotropic solids, the three-dimensional (3D) direction-dependent Young's modulus, linear compressibility, shear modulus, and Poisson's ratio should all have spherical forms. Any deviation from spherical shape, on the other hand, demonstrates anisotropy. For the $Mo_3P$ compound, we have displayed the ELATE [55] generated 2D and 3D plots showing the directional dependences of Young's modulus, shear modulus, linear compressibility, and Poisson's ratio.

As can be observed in Figures 2 to 5, there is a slight deviation from the spherical shape in the 3D figures of Y, G, and σ, indicating some anisotropy. However, there is no deviation from spherical shape in the 3D figure of compressibility. Therefore, compressibility and bulk modulus of $Mo_3P$ is isotropic. The ELATE plots also show that the projections of the direction dependent Y, G, and σ in the *ab*-plane are reasonably circular. This means that the elastic anisotropy within the basal plane is extremely low.

**3.2 Phonon dynamics**

The properties of phonons are extremely important in solid state physics. Phonon dynamics influence all thermal and charge transport parameters. The nature of the phonon modes governs the dynamical and structural stabilities of a crystal. Many features of a material can be measured directly or indirectly using phonon dispersion spectra (PDS) and phonon density of states (PHDOS) [56]. Phonon dispersion curves provide information regarding a material's dynamical stability, structural phase transition, and contribution of atomic vibrations to its thermal properties [57]. In this work, to test the dynamical stability, the phonon dispersion curves and phonon density of states of the $Mo_3P$ compound at absolute zero were determined.

This computation was carried out using the finite displacement method (FDM) based on density functional perturbation theory (DFPT) [20,21]. Figure 6 depicts the phonon dispersion spectra and phonon density of states of the mentioned materials at different pressures along the high symmetry direction of the BZ.

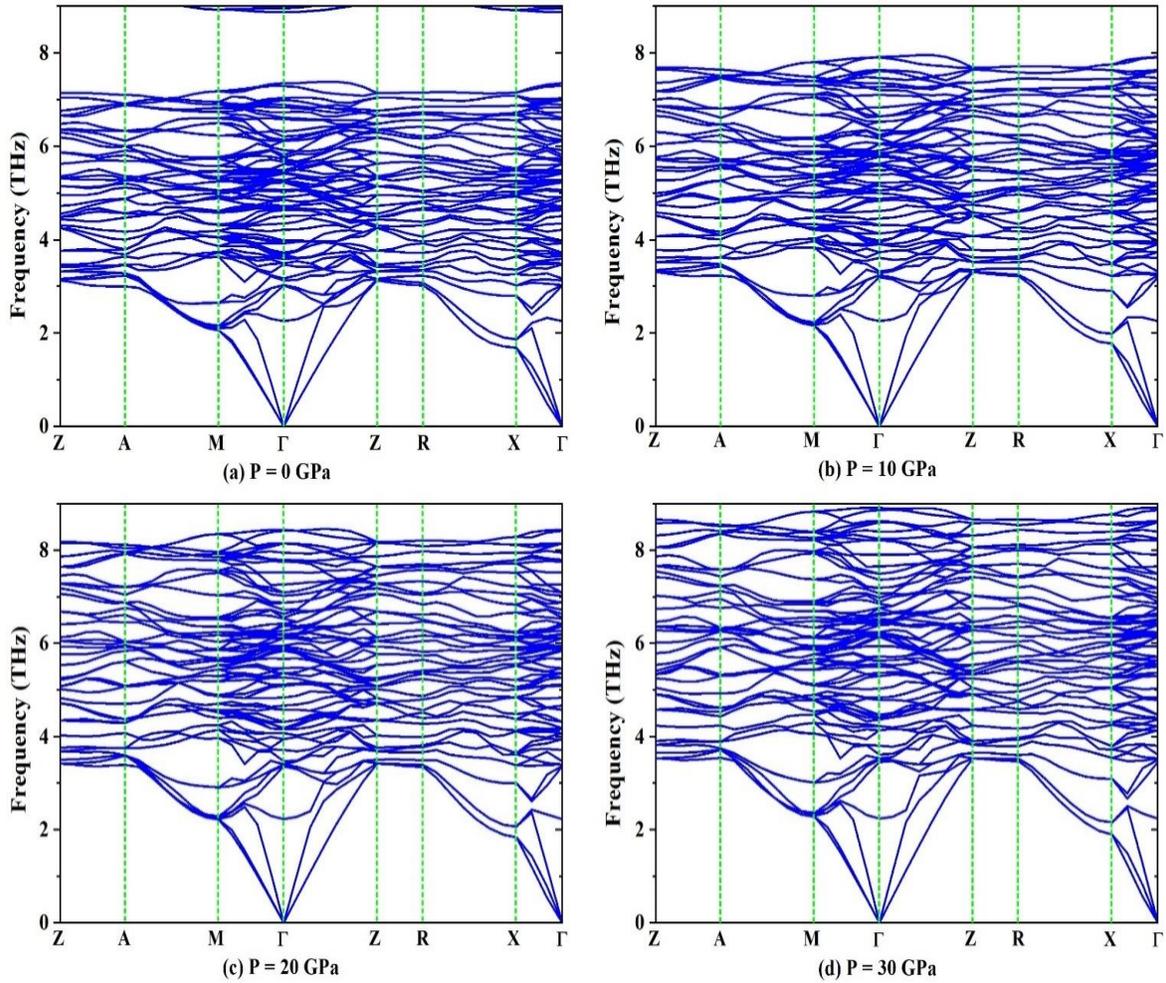

**Figure 6.** Phonon dispersion curves of $Mo_3P$ for (a) P = 0 GPa, (b) P = 10 GPa, (c) P = 20 GPa and (d) P = 30 GPa.

The PDS show no negative frequency branch within the BS. This suggests that the compound is dynamically stable in the pressure range considered in this study.

There are three phonon branches in the low frequency. These are the acoustic modes. These modes are divided into transverse acoustic (TA) and longitudinal acoustic (LA) modes. The branches in the high frequency, on the other hand, are optical modes. In general, low-frequency acoustic modes are caused by the vibration of heavy atoms, whereas high-frequency optical modes are caused by the out of phase vibrations of light atoms. The pressure dependent PHDOS profiles are shown in Figure 7. The $Mo_3P$ compound does not have a phonon gap between acoustic and optical modes. The high frequency optical modes are less dispersive.

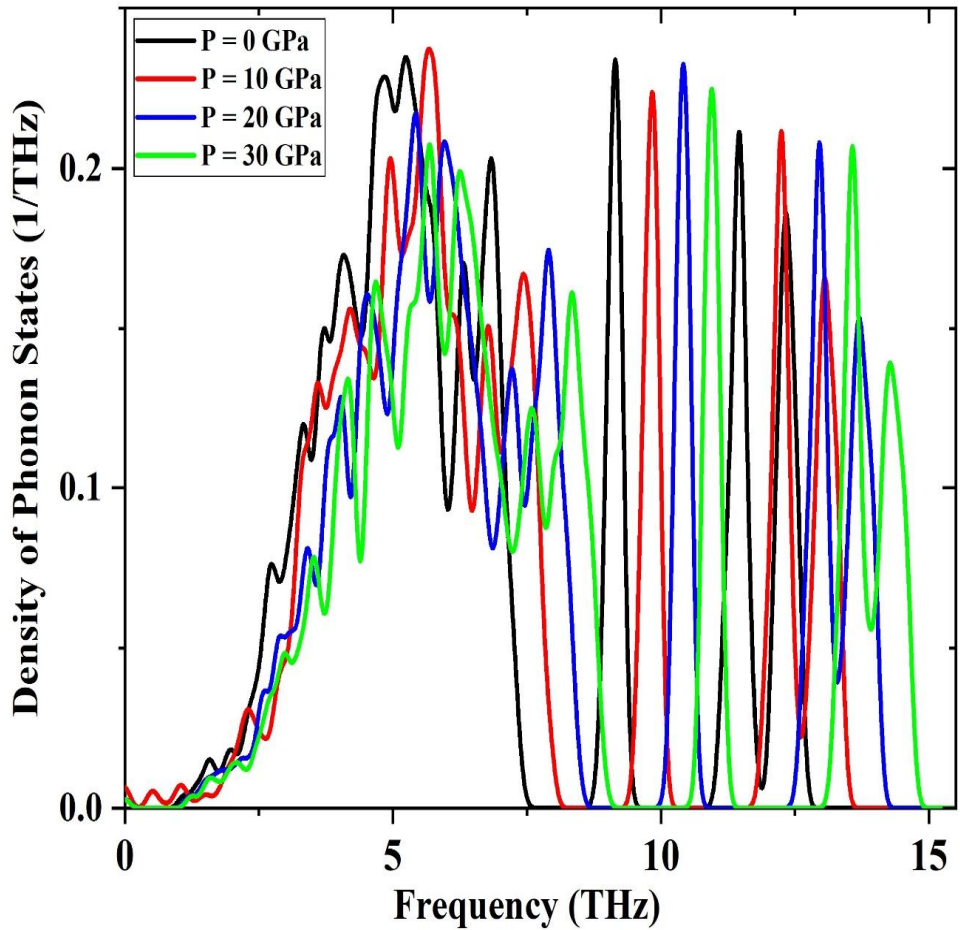

**Figure 7.** Phonon density of states (PHDOS) of $Mo_3P$ for various pressures.

These branches should lead to high PHDOS peaks as seen in Fig. 7. Application of pressure does not affect the nature of dispersion to any large extent. On the other hand, various phonon branches shift to higher frequencies as pressure increases. This effect is also seen in the positions of the PHDOS peaks shown in Figure 7. The reason of this shift is the pressure induced stiffening of the crystal which increases the effective stiffness constant of Mo$_3$P.

### 3.3. Electronic band structure and density of states

Figure 8 depicts electronic band structures for the optimized crystal structure of Mo$_3$P along different high symmetry axes (*X-R-M-Γ-R*) in the first Brillouin zone. The Fermi level is indicated by the horizontal broken line. The band structure characteristics clearly show that several bands with varied degrees of dispersion cross the Fermi level. This demonstrates the metallic nature of the Mo$_3$P at different pressures.

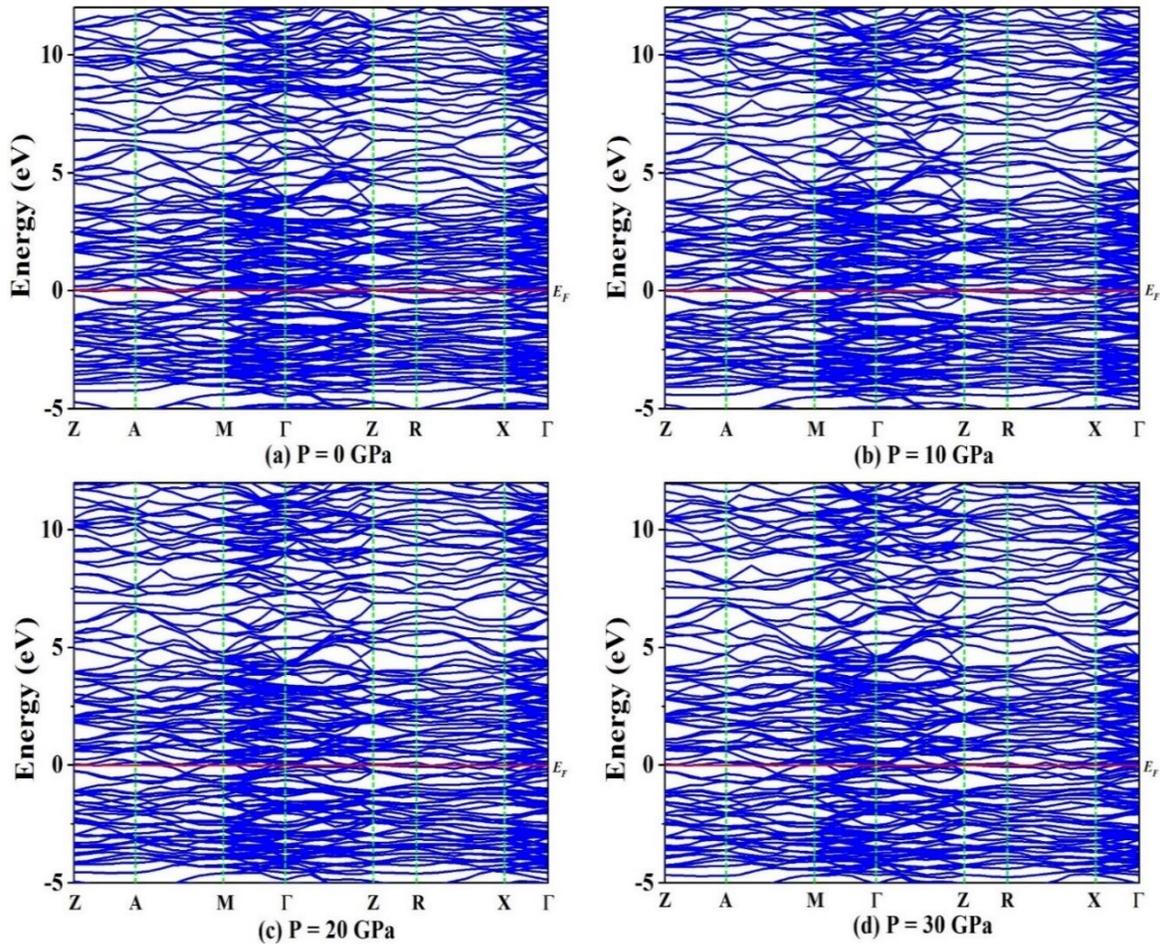

**Figure 8.** Electronic band structure of Mo$_3$P for (a) P = 0 GPa, (b) P = 10 GPa, (c) P = 20 GPa and (d) P = 30 GPa, along the high symmetry directions of the *k*-space within the first Brillouin zone.

The bands crossing the Fermi level show varying degree of dispersion at various regions of the *k*-space. Both electron- and hole-like bands cross the Fermi level. The effect of pressure is moderate on the band structure. Highly dispersive bands imply low charge carrier effective mass and high charge mobility. The situation becomes opposite for non-dispersive bands.

The calculated total and partial density of states (TDOSs and PDOSs, respectively) of $Mo_3P$ compound, as a function of energy, ($E$-$E_F$), are represented in Figure 9 for different pressures. The vertical dashed line at 0 eV marks the Fermi level, $E_F$. To understand the contribution of each atom to the TDOSs, we have calculated the PDOSs of Mo and P atoms in $Mo_3P$. The non-zero values of TDOSs at the Fermi level is the evidence that $Mo_3P$ should exhibit metallic electrical conductivity. At the Fermi energy, the values of TDOSs for $Mo_3P$ are 22.7, 21.9, 20.9 and 20.2 states per eV per unit cell at pressures of 0, 10, 20 and 30 GPa, respectively. The TDOS decreases gradually with increasing pressure. Hence one can say that $Mo_3P$ should be the most conducting at the low pressure. Near the Fermi level, the main contribution to TDOSs comes from the Mo-4*d* orbitals and P-3*p* orbitals of $Mo_3P$. Thus, these electronic states dominate the electrical conductivity and other transport coefficients of $Mo_3P$. The peaks in the TDOS controls the energy dependent optoelectronic properties of the compound to a large extent.

The structural, electronic and mechanical stabilities of $Mo_3P$ are affected by the properties of Mo-4*d* and P-3*p* orbitals bonding electronic states. It seen from the PDOS that there is a significant overlap in energy between the Mo-4*d* and P-3*p* bands in $Mo_3P$ compound. Such overlaps are suggestive of bonding between the electronic states involved. The nearest peak at the negative energy below the Fermi level in TDOS is known as the bonding peak, while the nearest peak at the positive energy is the anti-bonding peak. The energy gap between these peaks is called the pseudo-gap which is an indication of electronic stability [58–60]. The electronic density of states profiles show that the peaks below the Fermi level in the valence band are red shifted and the peaks above the Fermi level in the conduction band are blue shifted with increasing pressure.

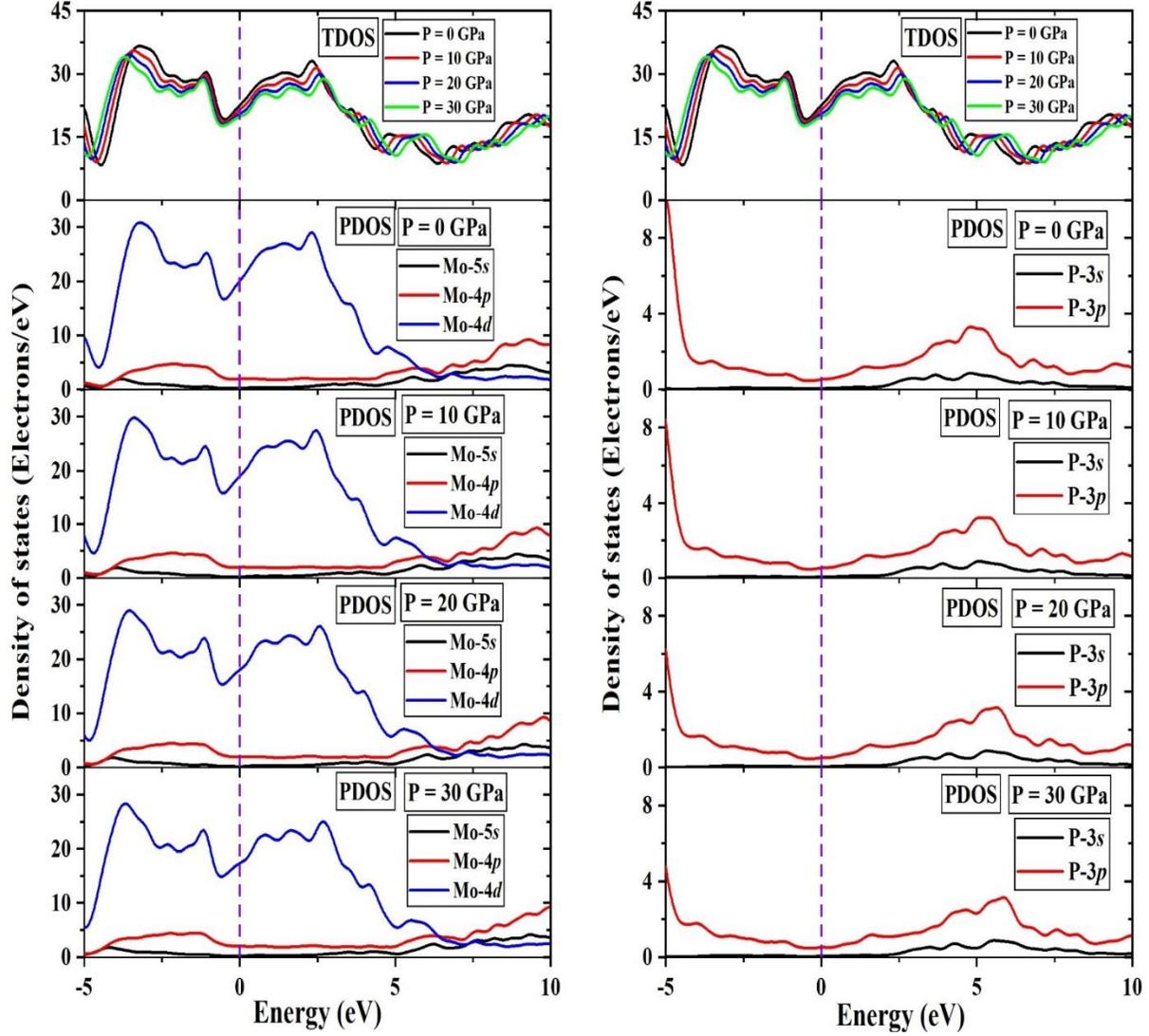

**Figure 9.** Total and partial electronic energy density of states (TDOS and PDOS) of Mo$_3$P at different pressures.

### 3.4 Thermophysical properties

**(a) Debye temperature and sound velocities**

The Debye temperature ($\Theta_D$) is a crucial thermophysical parameter of solids. It is linked to a variety of physical properties e.g., lattice vibration, thermal conductivity, phonon specific heat, interatomic bonding, vacancy formation energy, melting temperature, and coefficient of thermal expansion. It also sets an energy scale for Cooper pairing in phonon mediated superconductors. The Debye temperature can be defined as the temperature at which all the atomic modes of vibrations in a solid become activated. The Debye temperature of the compound depends on the crystal stiffness and atomic masses of its constituent atoms. In general, solids with stronger interatomic bonding strength, higher melting temperatures, greater hardness, higher acoustic wave velocity, and lower average atomic mass are shown to have larger Debye temperatures. Moreover, the Debye temperature marks a boundary between the classical and quantum behavior in lattice vibration. When $T > \Theta_D$, all the modes of vibrations get an energy $\sim k_B T$. On the other hand, when $T < \Theta_D$, the higher frequency modes are considered to be frozen and the quantum mechanical nature of vibrational modes are manifested [61]. The Debye temperature can be obtained using the following equation [62,63]:

$$\Theta_D = \frac{h}{k_B}\left(\frac{3N}{4\pi V_O}\right)^{\frac{1}{3}} v_a \qquad (40)$$

In the above expression, $h$ is the Planck's constant, $k_B$ is the Boltzmann's constant, $n$ is the number of atoms in the unit cell, $M$ is the molar mass, $\rho$ is the density of the solid, $N_A$ is the Avogadro number, and $v_a$ denotes mean sound velocity. The mean value $(v_a)$ can be calculated from bulk the ($B$) and shear ($G$) moduli using the longitudinal ($v_l$) and the transverse ($v_t$) sound velocities as follows [62]:

$$v_t = \sqrt{\frac{G}{\rho}} \quad and \quad v_l = \sqrt{\frac{B+\frac{4G}{3}}{\rho}} \qquad (41)$$

The transverse and longitudinal sound velocities can be used to calculate the average sound velocity, $v_a$ by using the following equation [47]:

$$v_a = \left[\frac{1}{3}\left(\frac{2}{v_t^3} + \frac{1}{v_l^3}\right)\right]^{-\frac{1}{3}} \qquad (42)$$

The calculated values of the Debye temperature, $\Theta_D$ and $v_l$, $v_t$, and $v_a$ at various pressures are given in Table 8. The calculated Debye temperature of Mo$_3$P is 466.44 K at 0 GPa. The value of Debye temperature increases gradually with increasing pressure.

The acoustic impedance of a material is a useful parameter for sound related applications. This parameter can be defined as:

$$Z = \sqrt{\rho G} \qquad (43)$$

The value of acoustic impedance of the compound $Mo_3P$ is $31.20 \times 10^6$ Rayl and increases with pressure.

The intensity or sound radiation factor is given by,

$$I = \sqrt{\frac{G}{\rho^3}} \qquad (44)$$

The calculated value of the sound radiation factor 0.40 $m^4$/kg.s for $Mo_3P$ at 0 GPa and this remains almost unchanged with changing pressure.

**Table 8.** Calculated mass density $\rho$ (kg/$m^3$), longitudinal velocity of sound $v_l$ ($ms^{-1}$), transverse velocity of sound $v_t$ ($ms^{-1}$), average sound velocity $v_a$ ($ms^{-1}$), Debye temperature $\Theta_D$ (K), acoustic impedance Z (Rayl) and radiation factor I ($m^4$/kg.s) of $Mo_3P$.

| Pressure, P (GPa) | $\rho$ | $v_l$ | $v_t$ | $v_a$ | $\Theta_D$ | Z (x$10^6$) | I |
|---|---|---|---|---|---|---|---|
| 0 | 9131.99 | 6612.10 | 3411.70 | 3820.00 | 466.44 | 31.20 | 0.40 |
| 10 | 9465.31 | 6979.96 | 3559.71 | 3988.56 | 492.93 | 33.69 | 0.38 |
| 20 | 9764.69 | 7290.75 | 3679.60 | 4125.52 | 515.18 | 35.93 | 0.38 |
| 30 | 10038.16 | 7564.49 | 3788.10 | 4249.16 | 535.52 | 38.03 | 0.38 |

Sound velocity in crystalline solid controls a large number of transport properties and the dynamical stability. For crystals with tetragonal symmetry, the pure transverse and longitudinal modes can only exist for the symmetry directions of type [010] (or [100]), [001] and [110]. Therefore, the acoustic velocities along these principal directions can be expressed as [24,63]:

[010] = [100]:

$$[010]v_l = [100]v_l = \sqrt{\frac{C_{11}}{\rho}} \; ; \; [001]v_{t1} = \sqrt{\frac{C_{44}}{\rho}} \; ; \; [010]v_{t2} = \sqrt{\frac{C_{66}}{\rho}} \qquad (45)$$

[001]:

$$[001]v_l = \sqrt{\frac{C_{33}}{\rho}} \; ; \; [100]v_{t1} = [010]v_{t2} = \sqrt{\frac{C_{66}}{\rho}} \tag{46}$$

[110]:

$$[110]v_l = \sqrt{\frac{(C_{11}+C_{12}+2C_{66})}{2\rho}} \; ; \; [001]v_{t1} = \sqrt{\frac{C_{44}}{\rho}} \; ; \; [1\bar{1}0]v_{t2} = \sqrt{\frac{(C_{11}-C_{12})}{2\rho}} \tag{47}$$

The values of these sound velocities (in ms$^{-1}$) of Mo$_3$P along different crystallographic directions are listed in **Table 9.**

**Table 9.** Sound velocities of Mo$_3$P (in ms$^{-1}$) along different directions.

| Propagation directions | | Anisotropic sound velocities (ms$^{-1}$) at pressures | | | |
|---|---|---|---|---|---|
| | | 0 GPa | 10 GPa | 20 GPa | 30 GPa |
| [100] = [010] | [100]v$_l$ = [010]v$_l$ | 6487.1 | 6833.7 | 7127.7 | 7391.0 |
| | [001]v$_{t1}$ | 3282.3 | 3417.6 | 3532.6 | 3633.7 |
| | [010]v$_{t2}$ | 3805.6 | 3999.3 | 4161.6 | 4301.3 |
| [001] | [001]v$_l$ | 6781.8 | 7192.7 | 7521.3 | 7818.6 |
| | [100]v$_{t1}$ | 3805.6 | 3999.3 | 4161.6 | 4301.3 |
| | [010]v$_{t2}$ | 3805.6 | 3999.3 | 4161.6 | 4301.3 |
| [110] | [110]v$_l$ | 6843.0 | 7239.8 | 7576.2 | 7874.0 |
| | [001]v$_{t1}$ | 3282.3 | 3417.6 | 3532.6 | 3633.7 |
| | [1$\bar{1}$0]v$_{t2}$ | 3120.6 | 3206.2 | 3274.9 | 3336.0 |

Sound velocity increases systematically with increasing pressure.

**(b) Melting temperature**

Another significant parameter for solid is its melting temperature ($T_m$) that gives the idea about the limit of temperature up to which the solid can be used. The melting temperature of a crystalline material is principally determined by its bonding energy and atomic arrangement. A material with a high value of melting temperature shows strong atomic bonding, a high value of bonding energy and a low value of thermal expansion coefficient [64]. We have calculated melting temperature $T_m$ using the following equation [65]:

$$T_m = 354 + 1.5(2C_{11} + C_{33}) \tag{48}$$

The calculated melting temperature of Mo$_3$P is 2137.05 K at 0 GPa. The value of melting temperature of Mo$_3$P is high and increases gradually with increasing pressure.

**(c) Thermal expansion coefficient and heat capacity**

The thermal expansion coefficient ($\alpha$), a material's intrinsic thermal property that is linked to a number of other physical properties such as specific heat, thermal conductivity, temperature variation of the energy band gap in semiconductors, and electron/hole effective mass, is crucial for the epitaxial growth of thin films. The thermal expansion coefficient (TEC), $\alpha$ is given by [24]:

$$\alpha = \frac{1.6 \times 10^{-3}}{G} \tag{49}$$

In this equation G is the shear modulus (in GPa). The calculated value of thermal expansion coefficient of Mo$_3$P is $1.50 \times 10^{-5}\ K^{-1}$ at 0 GPa. From Table 10, it was seen that the value of thermal expansion coefficient decreases gradually with increasing pressure.

**(d) Dominant phonon mode**

Phonons or quantum of lattice vibrations of materials make a significant contribution to a number of physical properties, such as electrical conductivity, thermopower, thermal conductivity and heat capacity. The dominant phonon wavelength of a compound, λ$_{dom}$, is the wavelength at which the phonon distribution function exhibits the peak. The wavelength of the dominant phonon at 300 K has been estimated by using the following expression [63,66]:

$$\lambda_{dom} = \frac{12.566 v_a}{T} \times 10^{-12} \qquad (50)$$

where, T is the temperature in degree K and $v_a$ is the average sound velocity in ms$^{-1}$. Materials with higher average sound velocity, higher shear modulus, and lower density exhibits higher wavelength of the dominant phonon mode. The obtained values of $\lambda_{dom}$ (in meter) of Mo$_3$P at various pressures are given in Table 10.

**Table 10.** Number of atoms per unit volume n (atoms/m$^3$), melting temperature $T_m$ (K), thermal expansion coefficient α (K$^{-1}$), and wavelength of dominant photon $\lambda_{dom}$ (m) at 300 K of Mo$_3$P.

| Pressure, P (GPa) | n (10$^{28}$) | $T_m$ | α (10$^{-5}$) | $\lambda_{dom}$ (10$^{-12}$) |
|---|---|---|---|---|
| 0 | 6.90 | 2137.05 | 1.50 | 159.99 |
| 10 | 7.15 | 2414.62 | 1.33 | 167.07 |
| 20 | 7.38 | 2670.85 | 1.21 | 172.80 |
| 30 | 7.58 | 2919.53 | 1.11 | 177.98 |

The dominant phonon mode shifts to lower wavelength as the pressure increases. This is a consequence of the pressure induced stiffening of the crystal. This finding also agrees completely with the lattice dynamical calculations.

**(e) Lattice thermal conductivity**

Lattice vibrations are an effective avenue of heat transport. The ability of a material to transport heat due to a temperature gradient is measured by the parameter known as the thermal conductivity. The lattice thermal conductivity ($k_{ph}$) of a material at different temperatures determines the amount of heat energy transported by the lattice vibration when there is a temperature gradient in the system. The $k_{ph}$ as a function of temperature can be calculated from an expression proposed by Slack [67] given below:

$$k_{ph} = A \frac{M_{av} \Theta_D^3 \delta}{\gamma^2 n^{2/3} T} \qquad (51)$$

where, $M_{av}$ is the average atomic mass in kg/mol, $\Theta_D$ is the Debye temperature in K, $\delta$ is the cubic root of the average atomic volume in meter, $\gamma$ refers to the acoustic Grüneisen parameter, $n$ is the number of atoms in the conventional unit cell, and $T$ is the absolute temperature in K. The Grüneisen parameter measures the degree of anharmonicity of the phonons. The acoustic Grüneisen parameter, $\gamma$ and the factor $A(\gamma)$ due to Julian [67], can be calculated by using the following relations [67,68]:

$$\gamma = \frac{3(1+s)}{2(2-3s)} \tag{52}$$

and,

$$A(\gamma) = \frac{5.720 \times 10^7 \times 0.849}{2 \times (1 - 0.514/\gamma + 0.228/\gamma^2)} \tag{53}$$

where, $\sigma$ refers to the Poisson's ratio. The lattice thermal conductivity is calculated at 300 K for all the pressures considered and the values obtained are given in Table 11. It is clearly seen from Table 11, that the values of $k_{ph}$ are increasing as the pressure increases. This is expected, since solids with higher Debye temperature have higher lattice thermal conductivities, as found in the preceding section. The overall thermal conductivity of this compound is high. Therefore, this material can be used as a heat sink.

**(f) Minimum thermal conductivity**

The behavior of a solid at temperatures higher than the Debye temperature has become an issue for high temperature applications. At high temperatures, a compound's inherent lattice thermal conductivity approaches a lowest value which is called the minimum thermal conductivity ($k_{min}$). This parameter is important for the fact that it is not affected by the existence of defects, impurities or imperfections in the crystal. Using the quasi-harmonic Debye model, Clarke developed the following equation for estimating the $k_{min}$ at high temperature [66]:

$$k_{min} = k_B \upsilon_a (V_{atomic})^{-\frac{2}{3}} \tag{54}$$

In the above expression, $V_{atomic}$ is the cell volume per atom of the compound.

The values of minimum thermal conductivity calculated from Clarke formula is given in Table 11. Again, we know that elastically anisotropic materials have different values of minimum thermal conductivity in different directions, i.e., they naturally exhibit anisotropy in minimum thermal conductivity. This anisotropy can be determined using the longitudinal and transverse sound speeds along different crystallographic axes. In this work, we used the following formula to compute the $k_{min}$ along various directions which is predicted by the Cahill model [64]:

$$k_{min} = \frac{k_B}{2.48} n^{2/3} (\upsilon_l + \upsilon_{t1} + \upsilon_{t2}) \qquad (55)$$

The calculated values of $k_{min}$ along the [100], [110] and [001] crystallographic directions are listed in Table 11.

**Table 11.** Grüneisen parameter γ, minimum thermal conductivities $k_{min}$ in different crystallographic directions (all in W/m-K), minimum thermal conductivity $k_{min}$ in Cahill's and Clark's method (both in W/m-K) and lattice thermal conductivity $k_{ph}$ (W/m-K) at 300 K of $Mo_3P$.

| Pressure, P (GPa) | γ | [100]$K_{min}$ or [010]$K_{min}$ | [001]$K_{min}$ | [110]$K_{min}$ | $K_{min}$ Cahill | $K_{min}$ Clark | $K_{Ph}$ |
|---|---|---|---|---|---|---|---|
| 0 | 1.89 | 1.27 | 1.35 | 1.24 | 1.29 | 0.89 | 0.70 |
| 10 | 1.93 | 1.37 | 1.46 | 1.33 | 1.39 | 0.95 | 0.78 |
| 20 | 1.97 | 1.45 | 1.55 | 1.41 | 1.47 | 1.00 | 0.84 |
| 30 | 1.99 | 1.53 | 1.64 | 1.48 | 1.55 | 1.05 | 0.91 |

From Table 11, it is seen that the Clarke model predicts a lower minimum thermal conductivity than the Cahill model. It is also observed that the $k_{min}$ of $Mo_3P$ compound is comparatively higher at higher pressures. It can also be concluded from the above table that the $k_{min}$ along [001] direction is slightly higher than that along [100] direction. This implies the existence of small anisotropy in the thermal transport. The minimum thermal conductivity of $Mo_3P$ is also significant.

## 3.5 Charge density distribution

Electron charge density distribution gives detailed information about a material's chemical bonding and shows charge transfer among the atoms. In this work, we have investigated the electronic charge density distribution of $Mo_3P$ at various pressures. Figures 10 and 11 depict the electronic charge density distribution at 0 GPa and 30 GPa, respectively. The color scale between the charge density maps shows the total electron density. The red color on the color scale represents the highest electron density whereas the blue color represents the lowest electron density.

It is clearly seen from Figure 10 that the charge distribution around Mo and P atom are almost spherical and there is high electron density around the Mo atom and low electron density around the P atom. Thus, ionic bonding dominates between these two atoms. But the slight deviation from the spherical shape and accumulation of small amount of charge in between the P-P and Mo-P atoms indicate that weak covalent bonding is also present. Nearly uniform low density of electronic charge throughout the crystal plane gives rise to the metallic bonding. Therefore, $Mo_3P$ has mixed ionic, covalent, and metallic bondings, where the ionic contribution dominates. The effect of pressure on the bonding characteristics is small.

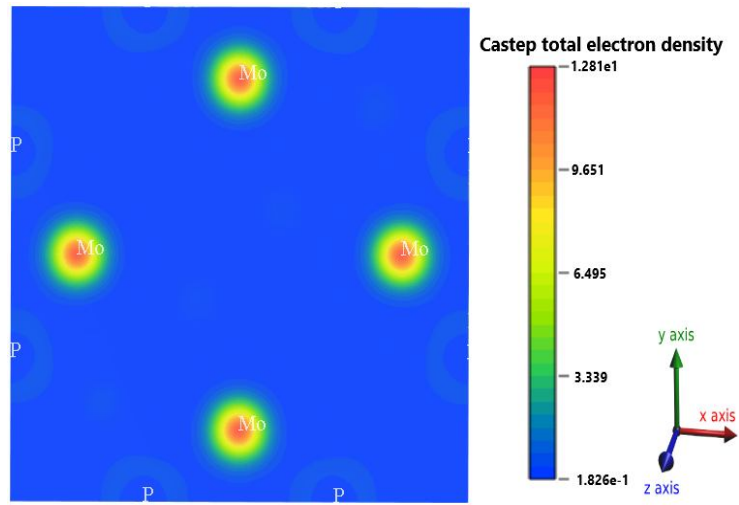

**Figure 10.** Electronic charge density map at 0 GPa.

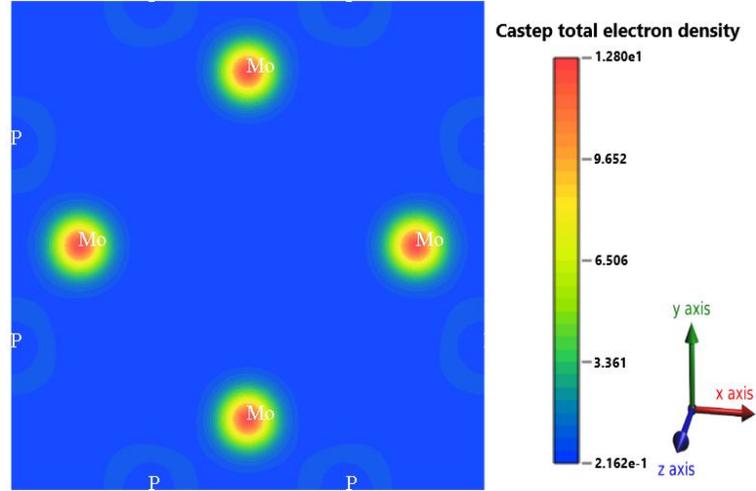

**Figure 11.** Electronic charge density map at 30 GPa.

### 3.6 Optical properties

A material's optical response can be determined by a number of energy/frequency dependent parameters, which include the absorption coefficient $\alpha(\omega)$, dielectric function $\varepsilon(\omega)$, energy loss function $L(\omega)$, optical conductivity $\sigma(\omega)$, refractive index $n(\omega)$, and reflectivity $R(\omega)$ [69]. These optical parameters are computed for pressures 0 GPa to 30 GPa and are presented in Figure 12. The optical spectra are shown for incident photon energies up to 30 eV in [100] polarization direction of the electric field vector.

The values of the $\alpha(\omega)$ can be used to understand a material's electrical nature. The $\alpha(\omega)$ of the $Mo_3P$ compound is shown in Figure 12. It is observed that photon absorption starts right from zero energy. This indicates that there is no optical band gap in $Mo_3P$. This also proves that the $Mo_3P$ is a conductor. Interestingly, $Mo_3P$ shows notable absorption in the mid-ultraviolet region of the electromagnetic spectrum. Also, the value of absorption coefficient $\alpha(\omega)$ increases slightly due to the increase of pressure.

The optical conductivity, $\sigma(\omega)$ is the electrical conductivity across a specified range of photon energy. Figure 12 shows that the $\sigma(\omega)$ of $Mo_3P$ material is starting from zero energy. It agrees with the previously computed electrical properties, implying that the material is a conductor once more. The optical conductivity decreases with increasing energy in the mid-ultraviolet (UV) region. At low energies, the optical conductivity is high for low pressures; the behavior changes in the UV region where the conductivity increases a little as the pressure increases.

Complex dielectric function is a vital optical parameter to describe a material's optical response since all the other energy dependent optical constants can be obtained from it. The variation of real (Re), $\varepsilon_1(\omega)$ and imaginary (Im), $\varepsilon_2(\omega)$ parts of the dielectric function of $Mo_3P$ is illustrated in Figure 12. From the figure, it is seen that the values of $\varepsilon_1(0)$ are almost same for all pressures. The imaginary part is related to dielectric loss. Figure 12 shows that the real part of dielectric function crosses zero from below at ~25 eV, whereas the imaginary part flattens to a very low value at the same energy indicating that the material will become transparent for photons with energies above 25 eV. This phenomenon confirms the Drude-like behavior (metallic feature) of $Mo_3P$.

The loss function, $L(\omega)$, of the $Mo_3P$ is shown in Figure 12. The position of the loss peak corresponds to the energy of the plasma excitations in the $Mo_3P$ compound. Loss function varies significantly with pressure and the heights of the loss peaks decrease with increasing pressure. At the same time, the peak energy shifts to higher values as pressure increases. This shift is mainly caused by the increase in the electron density in $Mo_3P$ with increasing pressure. The loss peak position coincides with the sharp falls in absorption coefficient and reflectivity as expected for materials.

Reflectivity, $R(\omega)$, measures the fraction of the incident light energy reflected from the material. The reflectivity spectra, $R(\omega)$, of $Mo_3P$ compound is depicted in Figure 12. It is found that $Mo_3P$ is a very good reflector of infrared and visible light. In the mid-UV region, reflectivity is lower but it increases again going through a pressure dependent peak in the energy range 20 eV to 24 eV. $R(\omega)$ falls sharply around 25 eV and $Mo_3P$ becomes transparent for energies over 25 eV to the incident electromagnetic wave.

The real part of the refractive index, $n(\omega)$, determines the group speed of an electromagnetic wave inside a material. The imaginary part (referred to as the extinction coefficient), $k(\omega)$, on the other hand, controls how much of the incident electromagnetic wave is attenuated when it passes through the substance. Figure 12 depicts the complex refractive indices of $Mo_3P$ crystal at different pressures. According to the plot, the value refractive index decreases with energy. Furthermore, it was seen that the value of refractive index of $Mo_3P$ is almost independent of pressure.

The overall results obtained in this section imply that there was no significant change on the optical properties of the compound Mo$_3$P due to the increase of pressure. This is consistent with the underlying electronic band structure which also shows weakly pressure dependent variations.

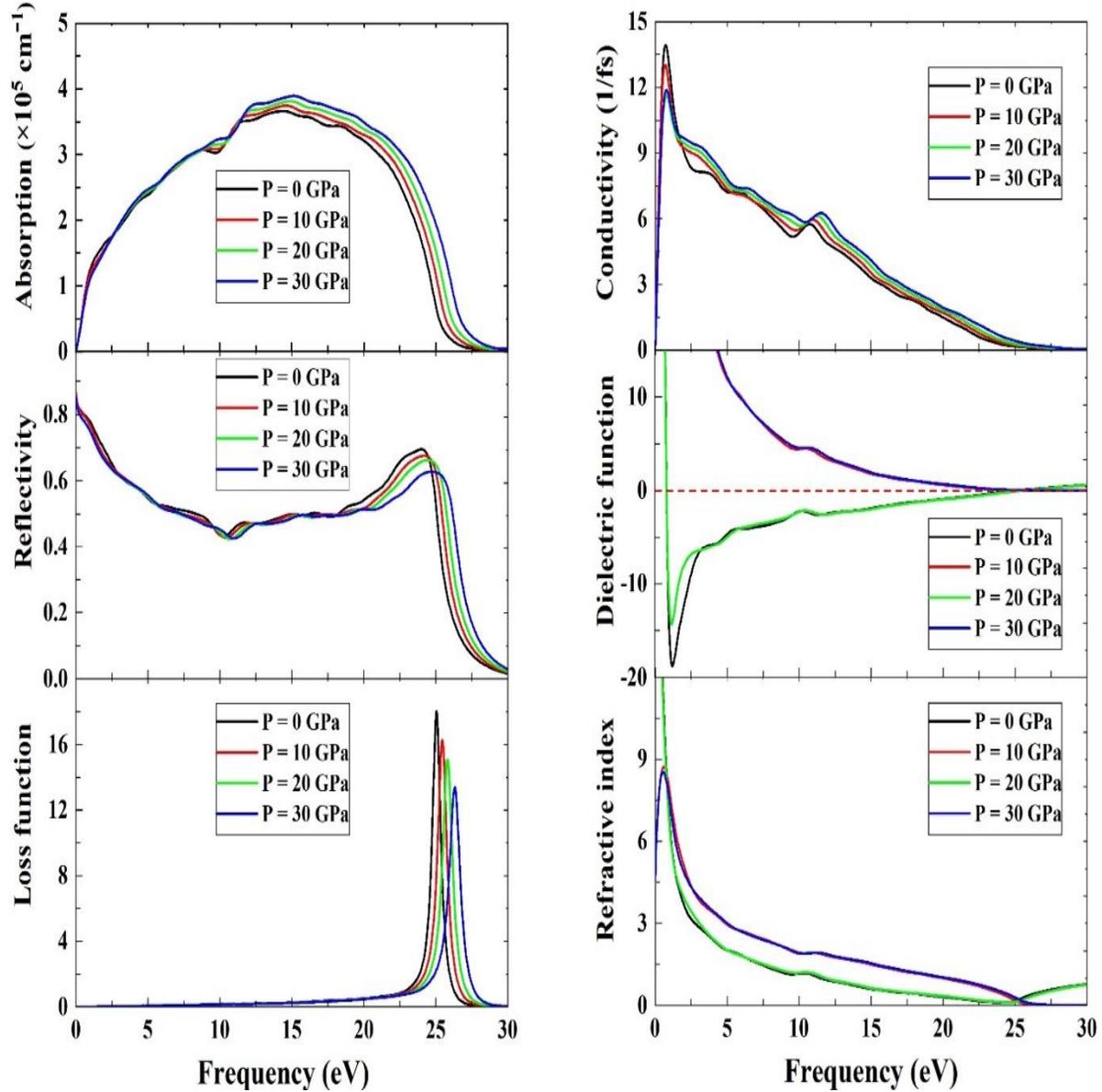

**Figure 12.** Optical properties (absorption coefficient, optical conductivity, reflectivity, complex dielectric function, loss function and complex refractive index) of Mo$_3$P compound at different pressures.

## 3.7 Superconducting state properties

Mo$_3$P exhibits phonon mediated superconductivity with a transition temperature, $T_c$, of 3.1 K [3]. The superconducting transition temperature of such systems can be estimated from the widely used expression proposed by McMillan [70]:

$$T_c = \frac{\theta_D}{1.45} exp\left[-\frac{1.04(1 + \lambda_{ep})}{\lambda_{ep} - \mu^*(1 + 0.62\lambda_{ep})}\right] \quad (56)$$

From the above equation, it is noted that the superconducting transition temperature depends on the Debye temperature, electron-phonon coupling constant ($\lambda_{ep}$), and the repulsive Coulomb pseudopotential ($\mu^*$). We observe that the Debye temperature of Mo$_3$P increases steadily with increasing pressure. The electronic density of states at the Fermi level [N(E$_F$)], on the other hand, decreases gradually with rising pressure. For a given average electron-phonon interaction energy, V$_{ep}$, the parameter $\lambda_{ep}$ varies linearly with N(E$_F$) via the relation, $\lambda_{ep}$ = N(E$_F$)V$_{ep}$ [71]. This shows that the electron-phonon coupling constant should decrease with increasing pressure in case of Mo$_3$P. The repulsive Coulomb pseudopotential, which works against the formation of Cooper pairs essential to superconductivity, also depends on the value of N(E$_F$) [72]. We have estimated this parameter for Mo$_3$P which decreases slightly as the pressure increases. Overall, the increase in the Debye temperature and the decrease in the Coulomb pseudopotential should favor an enhancement in the superconducting transition temperature with increasing pressure. But the decreasing trend in the pressure dependent $\lambda_{ep}$ should work in the opposite way. Thus, we predict a weak pressure dependent variation in $T_c$ for Mo$_3$P. For detailed and precise calculation of $T_c$, information regarding the Eliashberg spectral function is needed [11, 73] which falls outside the scope of CASTEP calculations.

## 4. Conclusions

In this paper, we have performed a detailed calculation of the structural, elastic, dynamical, optoelectronic and some thermophysical properties of non-centrosymmetric superconductor Mo$_3$P at various pressures using the first-principles approach based on DFT. The majority of the findings dealing with the properties of elastic, electrical, bonding, thermophysical, and optical properties are novel. The estimated lattice parameters correspond well with previously published data [7].

The calculated six independent elastic constants ($C_{11}$, $C_{12}$, $C_{13}$, $C_{33}$, $C_{44}$ and $C_{66}$) of Mo$_3$P compound satisfy the Born-Huang stability criteria of tetragonal crystals, suggesting that Mo$_3$P crystal is mechanically stable. According to Pugh's ratio (*G/B*), Poisson's ratio (*σ*) and Cauchy pressure (*C″*), Mo$_3$P compound is ductile in nature. The value of $\mu_M$ in Mo$_3$P implies a good level of machinability. Both ductility and the machinability increase with increasing pressure. Mo$_3$P compound shows excellent dry lubricating properties, lower feed forces, lower friction value and higher plastic strain value at high pressure. By hardness calculations of Mo$_3$P through different formulae, it is seen that this material shows reasonable hardness, and the hardness value increases with pressure. Ductility, machinability, and hardness values of Mo$_3$P suggest that this compound is suitable for engineering applications. In this respect, the physical properties of Mo$_3$P compares well with many of the technologically important MAX and MAB phase compounds [74-77]. The compound under study is moderately anisotropic. The electronic band structure and total density of states analysis show that Mo$_3$P exhibits clear metallic behavior. The calculated Debye temperature of Mo$_3$P increases almost linearly with pressure for the range considered. It is found that the melting temperature of Mo$_3$P compound is high which implies that Mo$_3$P is a good candidate material for high-temperature application. The phonon thermal conductivity of Mo$_3$P is high and it has potential to be used as a heat sink material. The optical constants' profiles of Mo$_3$P match very well with the electronic band structure and DOS profiles. Mo$_3$P compound is a good absorber of mid-ultraviolet radiation and is also an efficient reflector of infrared-visible light. The effect of pressure on the optical constants is low. By considering various parameters related to superconductivity, we qualitatively predict that the effect of pressure on superconductivity should be weak in case of Mo$_3$P.

**Declaration of interest**

The authors declare that they have no known competing financial interests or personal relationships that could have appeared to influence the work reported in this paper.

**Data availability**

The data sets generated and/or analyzed in this study are available from the corresponding author on reasonable request.

**CRediT author statement**

**Md. Sohel Rana:** Methodology, Software, Formal analysis. **Razu Ahmed:** Methodology, Software, Writing- Original draft. **Md. Sajidul Islam**: Software, Validation. **R.S. Islam:** Supervision, Writing- Reviewing and Editing. **S.H. Naqib:** Conceptualization, Supervision, Formal analysis, Writing- Reviewing and Editing.